\documentstyle[12pt]{article}
\newcommand{\fr}{\frac}
\newcommand{\lb}{\label}

\newcommand{\be}{\begin{equation}}
\newcommand{\ee}{\end{equation}}
\newcommand{\beqa}{\begin{eqnarray}}
\newcommand{\al}{\alpha}

\newcommand{\eeqa}{\end{eqnarray}}

\newcommand{\omo}{\omega}
\newcommand{\vp}{\varphi}
\begin{document}
\begin{flushright}
MRC.PH.-19/94

Revisted (will appear in J. Phys. A)6.
\end{flushright}

\vspace{1cm}
{\Large {\bf
\noindent
$sl_q(2)$ Realizations for Kepler
and Oscillator Potentials and
q-Canonical Transformations} }

\vspace{1.5cm}

\noindent
\"{O}. F. DAYI$^{a,}$\footnote{E-mail address:
dayi@yunus.mam.tubitak.gov.tr.},
I. H. DURU$^{a,b,}$\footnote{E-mail address:
duru@yunus.mam.tubitak.gov.tr.}

\vspace{1cm}

{\small {\it
\noindent
a) T\"{U}B\.{I}TAK-Marmara Research Centre,
Research Institute for Basic Sciences,
Department of Physics,
P.O.Box 21, 41470 Gebze, Turkey,

\noindent
b)Trakya University,
Mathematics Department,
P. O. Box 126, Edirne,  Turkey. } }

\vspace{4cm}

\date{}

{\small
\begin{center}
{\bf Abstract}
\end{center}

The realizations of the Lie algebra corresponding to the dynamical
symmetry group $SO(2,1)$ of the Kepler and oscillator
potentials are q-deformed.
The q-canonical transformation
connecting two realizations is given and a general  definition for
q-canonical transformation is deduced.
q-Schr\"{o}dinger equation
for a Kepler like potential
is obtained
from the q-oscillator Schr\"{o}dinger equation.
Energy spectrum
and the ground state wave function are calculated.}

\vspace{1cm}

\pagebreak

\section{Introduction}

There are mathematical and physical aspects
of q-deformations\cite{gen}. From the mathematical point
of view, one usually demands that the q-deformed algebra to be
a Hopf algebra. The physical point of view is somehow,
less restrictive: obtaining the underlying
undeformed picture in $q\rightarrow 1$ limit is the
basic condition.
Hence, q-deformation of a
physical system is not unique.
For example, the harmonic oscillator which is the most
extensively studied system, has several q-deformed
descriptions\cite{har}. q-deformation of physical systems
other than the oscillator are not well studied.
On the other hand, most of the
concepts of classical
and quantum mechanics become obscure after q-deformations.
For example, q-deformed change of phase space variables
leaving basic q-commutation relations invariant is
presented\cite{zu}, and a canonical transformation connecting
q-oscillators is studied\cite{fgig}; but, q-canonical transformations
establishing relationships between different potentials
are not known.

The purpose of this work is to present a q-canonical
transformation and to define a q-deformed Kepler
like potential in a consistent manner with
the q-oscillator. The posession of the same
dynamical symmetry group $SO(2,1)$ by the harmonic
oscillator and the Kepler problems will guide us.

In general, the phase space realizations of the
Lie algebras corresponding to a dynamical symmetry group
which are relavent to different physical systems are
connected by canonical transformations. We generalize this
connection to define q-canonical transformations.
We hope that the procedure may also help to define
new q-deformed potentials from the known ones.

In section 2 we review the known relation between
the undeformed Kepler and oscillator problems.

In section 3 we present  q-deformations of two  realizations
of $sl(2)$ (which is the Lie algebra
of $SO(2,1)$) relavent to the Kepler and oscillator
potentials.  We define the q-canonical transformation
connecting two realizations. We then give a general  definition of
q-canonical transformation.

In section 4 we define a q-Schr\"{o}dinger equation
for a Kepler like potential
from the q-Schr\"{o}dinger equation of the
q-oscillator by a coordinate change.

Finally, we
fix the energy spectrum of the q-oscillator,
and find the ground state wave function; then,
obtain  energy spectrum and ground state wave function of
the q-deformed Kepler problem.

\section{Review of the Relations between Kepler
and Oscillator Potentials}

It is well known that $SO(2,1)$ is the dynamical symmetry
group of the radial parts of the Schr\"{o}dinger equations
of the Kepler and the harmonic oscillator potentials\footnote{See
for example \cite{bar}.}.

In one (space) dimension
the phase space realizations of the corresponding
Lie algebra
$sl(2)$ relavent to the Kepler and the harmonic oscillator problems
are given by
\beqa
H & = & 2px, \nonumber \\
X_+ & = & -\sqrt{2}\ x ,\lb{fqa} \\
X_- & = & \fr{-1}{\sqrt{2}}p^2 x , \nonumber
\eeqa
and
\beqa
L_0 & = & up_u +\fr{i}{2}. \nonumber       \\
L_+ & = & -\sqrt{2}u^2, \lb{dgs} \\
L_- & = & -\fr{1}{4\sqrt{2}} p_u^2, \nonumber
\eeqa
with
\[
px-xp=i,\   p_uu-up_u=i.
\]

The above generators satisfy the usual commutation relations
\beqa
{[H,X_\pm ]}=\pm 2iX_\pm, & [X_+,X_-]=-iH, &  \\
& &  \nonumber \\
{[L_0,L_\pm ]}=\pm 2iL_\pm, & [L_+,L_-]=-iL_0. &   \lb{a2}
\eeqa

The eigenvalue equation for the Kepler Hamiltonian
\[
H_{K}\Psi \equiv \left( \fr{p^2}{2\mu }+
\fr{\beta^2}{x}\right) \Psi =E\Psi ,
\]
which is equivalent to
\[
\left( \fr{p^2}{2\mu }-E \right) x\Phi =\beta^2 \Phi ,
\]
is solved by diagonalizing the operator
\[
\fr{1}{\sqrt{2}} \left( \fr{1}{\mu} X_-+ EX_+ \right) .
\]

On the other hand solution of the oscillator problem is simply
obtained by diagonalizing the operator
\[
-\fr{1}{\sqrt{2}}\left(\fr{1}{\mu}L_-+\fr{1}{2}\mu \omo^2L_+ \right).
\]

Classicaly (i.e. before the $\hbar$-deformation)
the Kepler and the oscillator phase space variables
are connected by the canonical transformation
\be
\lb{ct}
x=u^2,\  p=\fr{p_u}{2u}.
\ee

The canonical transformations of the above type
are also employed for solving the H-atom path integral\cite{dk}.
In fact, since the path integrations make use
of the classical dynamical variables, the canonical
point transformations are rutinley used to transform
the path integral of a given potential into a
solvable form.

Relation between the Schr\"{o}dinger equations
corresponding to the one-dimensional
oscillator and Kepler type
potentials is the following.

Schr\"{o}dinger equation of the one-dimensional oscillator
\be
\lb{ose}
\left( -\fr{1}{2\mu}\fr{d^2}{du^2} +\fr{1}{2} \mu \omo^2 u^2
\right) \psi =E\psi ,
\ee
is transformed by the coordinate change suggested by
(\ref{ct})
\be
\lb{cov}
u=\sqrt{x},
\ee
into
\be
\lb{kse}
\left( -\fr{1}{2\mu}\fr{d^2}{dx^2} +
\fr{E/8}{x}-\fr{3/32\mu}{x^2}
\right) \phi =-\fr{\mu \omo^2}{8}\phi ,
\ee
with
\be
\lb{cov1}
\psi=\fr{1}{\sqrt{x}}\phi .
\ee
The energy $E$ and the frequency $\omo^2 $
of the oscillator play the role of the coupling constant
$\beta^2$ and the energy $E_K$  of the Kepler problem:
\beqa
\fr{E}{8}= & \fr{\omo}{4}(2n+1) & =-\beta^2, \lb{cac} \\
E_K = & -\mu \omo^2 /8 .  &   \lb{ken}
\eeqa

(\ref{kse}) is equivalent to the one-dimensional Kepler
problem with an extra potential barrier $-(3/32 \mu )/x^2$,
or to the two-dimensional Kepler problem with
``angular momentum" $p^2_\theta = -3/16.$

If we solve $\omo $ from (\ref{cac}) as
\be
\lb{os}
\omo (\beta ) =-\fr{4\beta^2}{2n+1},
\ee
and insert into (\ref{ken}), we obtain the
Kepler energy
\be
\lb{kee}
E_K=-\fr{2\mu\beta^4}{(2n+1)^2} .
\ee

\section{q-canonical Transformation between the Kepler
and Oscillator Realizations of $sl_q(2)$}

\subsection{q-deformation of the Kepler Realization}

To q-deform the algebra of the generators (\ref{fqa}),
we prefer to q-deform the commutation relation
between
$p$ and $x$
(we use the same notation for the q-deformed
and undeformed objects),

\be
\lb{fct}
xp -qpx=-i\sqrt{q},
\ee
but keep the functional forms of $H,X_\pm $ same as
the ones given in (\ref{fqa}).
The  q-deformed commutation relations are then given
by\cite{omer}

\pagebreak

\beqa
HX_- -qX_-H  &  =  & -2i\sqrt{q}X_-  \nonumber  \\
HX_+ -\fr{1}{q}X_+H  &  =  & 2i\fr{1}{\sqrt{q}} X_+  \lb{ffqa}  \\
X_+X_- -q^2 X_-X_+ &  =  &  \fr{-i}{2} \sqrt{q}(1+q) H .\nonumber
\eeqa
Note that after rescaling the above generators
\[
X_\pm \rightarrow \fr{i\sqrt{\sqrt{q}(1+q)}}{\sqrt{q}} X_\pm ,
\   H\rightarrow 2i H,
\]
and  setting $q=r^2$, one arrives at
\beqa
HX_- -r^2X_-H  &  =  & -rX_-  \nonumber  \\
r^2HX_+ -X_+H  &  =  & rX_+  \lb{qaw}  \\
X_+X_- -r^4 X_-X_+ &  =  & r^2 H ,\nonumber
\eeqa
which is the Witten's second deformation of $sl(2)\cite{wit}.$

\subsection{q-deformation of the Oscillator Realization}

We like to q-deform the generators given in (\ref{dgs}),
which are relavent to the oscillator problem,
in a consistent manner with the deformation of the
Kepler realization (\ref{fqa}).

We define the q-deformed commutation relation of $p_u$ and $u,$ as
\be
\lb{sct}
up_u -\sqrt{q}p_uu =ib(q),
\ee
and fix $b(q)$
by requiring the commutation relations of the q-deformed
algebra of the generators (\ref{dgs}) to be the same as (\ref{ffqa}).
We rescale $L_0,L_\pm ,$
\[
L_0\rightarrow \fr{a}{2\sqrt{q}} L_0,\    L_\pm
\rightarrow \left[\fr{a(1+\sqrt{q})^3}{8\sqrt{q}(1+q)}
\right]^{1/2} L_\pm ,
\]
with
\[
a=\fr{(1+\sqrt{q})(1+q)}{2\sqrt{q}},
\]
and fix $b(q)$ as
\[
b=-\fr{1}{2}(1+\fr{1}{ \sqrt{q}}).
\]
The q-deformed algebra, then becomes
\beqa
L_0L_- -qL_-L_0 &  =  & -2i\sqrt{q}L_-  \nonumber  \\
L_0L_+ -\fr{1}{q}L_+L_0 &  =  & 2i\fr{1}{\sqrt{q}} L_+  \lb{ggqa}  \\
L_+L_- -q^2 L_-L_+ &  =  &  \fr{-i}{2} \sqrt{q}(1+q) L_0 ,\nonumber
\eeqa
which is the same as the $sl_q(2)$ algebra of the
Kepler problem (\ref{ffqa}).

Note that before q-deformation, $sl(2)$ algebra (\ref{a2})
admits three different choices for
$L_0$
\be
\lb{od}
L_0=up_u + \fr{i}{2},\
L_0=p_u u - \fr{i}{2},\   L_0=\fr{1}{2}(up_u+p_uu).
\ee
In the q-deformed case however,
if we like to have the generators to be independent of q
(except an overall factor),
the ordering degeneracy in (\ref{od}) is removed,
that is $L_0$ can only take the  form given in
(\ref{dgs}).

\subsection{q-canonical Transformation}

Let us introduce a transformation similar to (\ref{ct}):
\be
\lb{qct}
x=\left( \fr{u}{b}\right)^2 =\left(
\fr{2\sqrt{q}}{1+\sqrt{q}}\right)^2 u^2,\
p=\fr{1}{2}u^{-1}p_u.
\ee
Then, the q-commutation relation (\ref{fct}) yields
\be
\lb{nqc}
\fr{q}{1+\sqrt{q}}(up_u -qu^{-1}p_uu^2) =-i\sqrt{q},
\ee
which is consistent with (\ref{sct}). Indeed,
by the virtue of (\ref{sct}), the above commutation
relation becomes
\[
\fr{q}{1+\sqrt{q}}(up_u-\sqrt{q}p_uu+ib\sqrt{q})=-i\sqrt{q},
\]
which is again equal to (\ref{sct}).

Now, we are ready to define q-canonical transformation.

{\bf Definition.} We like to keep the phase space realizations
of the q-deformed generators to be formally the same as the
undeformed  generators of the dynamical symmetry group.
We then define the transformation
$x,p\rightarrow u,p_u$ to be the q-deformed canonical transformation
if

\noindent
{\bf i)} algebras generated by the realizations
$X_i(x,p)$ and $L_i(u,p_u)$ are the same and,

\noindent
{\bf ii)} the q-commutation relations between
$p$ and $x,$ and $p_u$ and $u$ are preserved.

In accordance with the above definition,
we conclude that (\ref{qct}) is
a q-canonical transformation.

By rescaling the q-canonical variables $p,\ x$ and $p_u,\ u$ as
\[
(x,p) \rightarrow q^{-1/ 4}(x,p), \
(u,p_u) \rightarrow \sqrt{|b|/\sqrt{q}}\ (u,p_u),
\]
and setting
\[
q \rightarrow q^{-2} ,
\]
the q-commutators (\ref{fct}) and (\ref{nqc}) become
\beqa
\lb{eqc}
px-q^2xp & = & i, \\
\lb{eqc1}
p_uu-qup_u & = & i.
\eeqa
In the rest of the paper, these q-commutation relations
will be used.

There is another definition of q-deformed canonical
transformation\cite{fgig}: phase space coordinates are
transformed under the condition that the q-commutators
remain invariant. On the other hand, our condition in the
definition of q-canonical transformation is to obtain
in a suitable limit, the undeformed mappings connecting
different potentials which possess the same dynamical
symmetry. Thus our definition of q-canonical
transformation is dynamics dependent, i.e. the basic
q-commutators are potential dependent (\ref{eqc}-\ref{eqc1}).

\section{q-Canonical Transformation from
q-Oscillator Schr\"{o}dinger Equation to
q-Kepler Problem }

Introduce the
q-deformed derivative $D_q(u)$\cite{vil}
\be
\lb{qd}
D_q(u)f(u) \equiv \fr{f(u)-f(qu)}{(1-q)u} .
\ee
In terms of this definition one can show that
\be
\lb{cr}
D_q(u)\{ f(u)g(u) \} = D_q(u)f(u) g(u)+f(qu)D_q(u)g(u).
\ee
Since the q-deformed derivative $D_q(u)$ satifies
\[
D_q(u)u-quD_q(u) =1,
\]
we can set
\[
p_u=iD_q(u),
\]
which is consistent with (\ref{eqc1}).

In terms of this q-differential realization one can
obtain the q-deformed Schr\"{o}dinger equation for the
q-oscillator
\be
\lb{qdse1}
(-\fr{1}{2\mu} D_q^2(u) +\fr{\mu}{2}\omo^2_q c^2_q(u) -\fr{1}{2}E_q )
\psi_q(u) =0,
\ee
where
\be
\lb{co}
c_q(u)=\sqrt{q} u,\   \omo_q=\left[ {\omo}  \right]_q \equiv
\fr{1-q^\omo}{1-q}.
\ee
Obviously, the choice (\ref{co}) is not unique\footnote{For example
see \cite{ww} and the references given therein.}.
The conditions to be satisfied are
\[
\lim_{q\rightarrow 1}c_q (u) =u,\  \lim_{q\rightarrow 1}\omo_q=\omo .
\]

We adopt the change of variable
suggested by (\ref{qct})
\be
\lb{u}
u=\sqrt{x}.
\ee
The q-derivative $D_q(u)$ transforms as
\be
\lb{du}
D_q(u)=(1+q)\sqrt{x}D_{q^2}(x).
\ee
$D_{q^2}(x)$ satisfies
\be
\lb{dq2}
D_{q^2}(x)x- q^2xD_{q^2}(x)=1,
\ee
hence in accordance with (\ref{eqc})
it can be identified with $-ip.$
Therefore, (\ref{u}) and (\ref{du}) are equivalent to
the q-canonical transformation (\ref{qct}).

The q-Schr\"{o}dinger equation (\ref{qdse1}) then becomes
\be
\lb{se2}
\left[ -\fr{1}{2\mu} (1+q)^2 x D_{q^2}^2(x)
-\fr{1}{2\mu}(1+q) D_{q^2}(x)
+\fr{1}{2\mu} [ \omo ]_q^2 qx -\fr{1}{2}E_q \right]  \phi_q(x) =0,
\ee
with
\be
\lb{ph}
\phi_q(x) =\psi_q(\sqrt{x}).
\ee

To get rid of the term linear in $D_q(x)$ in (\ref{se2}),
we introduce the ansatz
\be
\lb{pp}
\phi_q(x)=x^\al \varphi_q(x).
\ee
Choosing
\[
\al=\fr{ {\rm ln}((3-q)/2)}{2 {\rm ln} q} ,
\]
and by multiplying (\ref{se2}) from left by $1/(1+q)^2x$, we
obtain
\beqa
\left[  \fr{-1}{2\mu}D^2_{q^2} (x)
-\fr{(2q^2 -2q -3)/ 8\mu q^2 (1+q)^2}{x^2}
-\fr{E_q/2 (1+q)^2}{x}
\right]  \vp_q(x) \nonumber \\
=\fr{\mu [ \omo ]_q^2q}{2(1+q)^2} \vp_q(x),
\eeqa
which is the q-deformed Schr\"{o}dinger equation of the
Kepler  potential with an extra potential barrier\footnote{
For a study of q-deformed H-atom in a unrelated manner to the
q-oscillator see \cite{qH}.}.

The q-oscillator energy $E_q$ is dependent on $\omo$
and q. From the identification of the coupling constant
\be
\lb{ka}
-\beta^2=E_q,
\ee
we can solve $\omo $ in terms of  $\beta$ and $q$,
as $\omo (\beta ,q).$
Hence in terms of the solutions of the q-Schr\"{o}dinger
equation for q-oscillator
(\ref{qdse1}) we can obtain the solutions of
\beqa
\left[ \fr{-1}{2\mu} D_{q^2}^2(x) + \fr{\beta^2/ 2(1+q)^2}{x}
+ \fr{(2q^2 -2q -3)/ 8\mu q^2 (1+q)^2}{x^2}  \right] \vp_q (x) \nonumber \\
=E_K \vp_q (x),
\eeqa
where
\be
\lb{enk}
E_K=\fr{q\mu [\omo (\beta ,q )]^2}{2(1+q)^2},
\ee
is the q-deformed anolog of the
energy spectrum of the Kepler problem.

\section{Energy Spectrum and Ground State Wave Functions}

General solution of the
q-Schr\"{o}dinger equation of the q-deformed
oscillator (\ref{qdse1}) is not known.
We fix the energy spectrum to be of the conventional
form\cite{har}
\be
\lb{eqn}
E_{qn}=[\omo (2n+1)]_q \equiv \fr{1-q^{\omo (2n+ 1)}}{1-q}.
\ee

Inserting the above energy spectrum into (\ref{ka})
we obtain
\be
\lb{wqn}
\omo_{qn}(\beta ,q)=\fr{1-[(1-q)(1+\beta^2)]^{{\rm ln} q / 2(2n+1)}}{1-q}.
\ee
The energy spectrum (\ref{enk}) of the q-Kepler problem then becomes
\be
E_{Kn}=\fr{q\mu}{2(1-q^2)^2}
\{1-[(1-q)(1+\beta^2)]^{{\rm ln} q / 2(2n+1)} \}^2.
\ee

We now like to build the ground state wave function of the q-oscillator.
For this aim, we introduce
\be
e_q(z^2) =1+\sum_{n=1}^\infty \left(
\prod_{k=1}^n
\fr{2(1-q)}{1-q^{2k}} \right) z^{2n},
\ee
which is defined to satisfy
\[
D_q(z)e_q(z^2) =2ze_q(z^2).
\]
Hence, the  equation for the ground state of the
q-oscillator
\be
\lb{qdse0}
(-\fr{1}{2\mu} D_q^2(u) +\fr{q}{2}\mu [\omo ]^2_q u^2)\psi_q^0(u) =
\fr{1}{2}[\omo ]_q
\psi^0_q(u),
\ee
possesses the solution
\be
\lb{gss}
\psi_q^0 (u) =e_q\left( (-\mu / 2)\fr{1-q^{\omo }}{1-q} u^2 \right) .
\ee

By introducing the above definition into (\ref{ph}) and
using (\ref{pp}) we obtain the ground state wave function
of the q-Kepler problem
\be
\vp_q^0 (x) =x^{-\al} \psi^0_q (\sqrt{x})=x^{-\al}
e_q\left( (-\mu / 2)\fr{1-q^{\omo }}{1-q} x \right) ,
\ee
which corresponds to the energy $E_{K0}$.

\pagebreak

\newcommand{\bi}{\bibitem}

\end{document}